\documentclass{aa}
\usepackage{epsf} 
\newcommand{\bg}[1]{\mbox{\boldmath $#1$}}
\def\lesssim{\mathrel{\hbox{\rlap{\hbox{\lower4pt\hbox{$\sim$}}}\hbox{$<$}}}}
\def\gtrsim{\mathrel{\hbox{\rlap{\hbox{\lower4pt\hbox{$\sim$}}}\hbox{$>$}}}}

\begin{document}

\thesaurus{02(02.19.2; 02.18.5; 02.16.1; 02.01.2)}

\title{Angular function for the Compton scattering in mildly and ultra
relativistic astrophysical plasmas}

\author{S.Y. Sazonov\inst{1,2}
\and R.A. Sunyaev\inst{1,2}}

\offprints{S.Y. Sazonov}

\institute{MPI f\"ur Astrophysik, Karl-Schwarzschild-Str.~1, 86740 Garching bei
M\"unchen, Germany
\and Space Research Institute (IKI), Profsoyuznaya~84/32, 117810
Moscow, Russia}

\date{Received December 10, 1999; accepted}

\titlerunning{Angular function for Compton scattering}
\authorrunning{Sazonov \& Sunyaev}

\maketitle

\begin{abstract}
Compton scattering of low-frequency radiation by an isotropic
distribution of (i) mildly and (ii) ultra relativistic electrons is
considered. It is shown that the ensemble-averaged differential 
cross-section in this case is noticeably different from the Rayleigh
phase function. The scattering by an ensemble of ultra-relativistic
electrons obeys the law $p=1-\cos{\alpha}$, where $\alpha$ is the
scattering angle; hence {\sl photons are preferentially scattered
backwards}. This contrasts the forward scattering behaviour in the
Klein-Nishina regime. Analytical formulae describing first-order
Klein-Nishina and finite-electron-energy corrections to the simple
relation above are given for various energy distributions of
electrons: monoenergetic, relativistic-Maxwellian, and power-law. A
similar formula is also given for the mildly relativistic (with
respect to the photon energy and electron temperature) corrections to
the Rayleigh angular function. One of manifestations of the phenomenon
under consideration is that hot plasma is more reflective with respect
to external low-frequency radiation than cold one, which is important, in
particular, for the photon exchange between cold accretion disks and
hot atmospheres (coronae or ADAF flows) in the vicinity of
relativistic compact objects; and for compact radiosources.

\keywords{scattering -- radiation mechanisms:
non-thermal -- plasmas -- accretion, accretion disks}

\end{abstract}

\section{Introduction}

The problem about the cross-section for scattering of radiation by
an individual free electron, i.e. Compton scattering, is classical and
its description can be found in any textbook on the interaction of matter
and light (see, e.g., Berestetskii et al. \cite{berestetskii}). The
Compton scattering of low-frequency radiation in a dense plasma for
which collective effects are important is also well studied. In
particular, formulae that describe the differential cross-section
averaged over an ensemble of plasma-bound electrons, or, stated
differently, the scattering angular function, are well-known (see,
e.g., Bekefi \cite{bekefi}; Zheleznyakov \cite{zheleznyakov}).  

\begin{figure}
\epsfxsize=9.5cm
\epsffile[85 275 560 595]{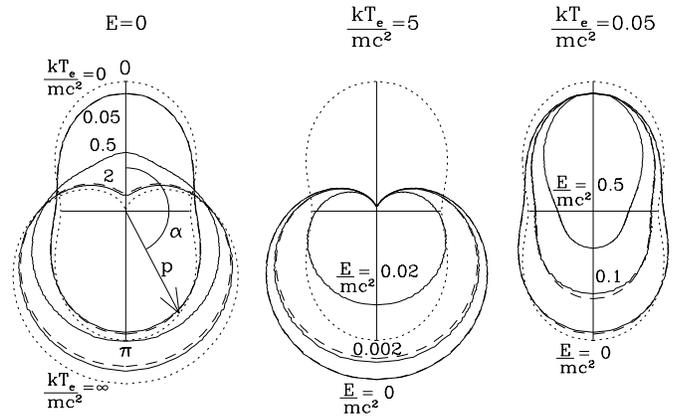}
\caption{Angular function (in polar coordinates) for the Compton
scattering of photons of energy $E$ by an ensemble of relativistic
Maxwellian electrons with temperature $T_e$. Solid lines show
Monte-Carlo simulation results. Dashed lines represent approximations
(when available) given either by Eq.~(\ref{ang_maxwel1}) or by
Eq.~(\ref{ang_maxwel2}). The dotted lines represent two extreme cases,
one of vanishing temperature, which corresponds to the Rayleigh
angular function, and the other of ultra-relativistic electrons
($kT_e\gg m_ec^2$). In the latter case (presented only in the left
panel), the scattering obeys the law $p=1-\cos{\alpha}$.
\label{figure}}
\end{figure}

In this Letter, we would like to point out the fact, which is
seemingly un-noticed in the literature, that the angular function for
Compton scattering can be different from the classical Rayleigh
function in the case of low-density relativistic electron gas too. We
show below, providing simple analytical formulae, that the differential
cross-section averaged over an isotropic ensemble of substantially
relativistic electrons (e.g., monoenergetic, Maxwellian, or with a
power-law energy spectrum) is backward-oriented, i.e. {\sl photons have a
tendency to be scattered backwards}, rather than forwards (Fig.~{\ref{figure}).
This phenomenon results from the combined action of two effects. One
is that a photon is more likely to suffer a scattering from an
electron that is moving toward it, rather than away from it (the
probability of scattering is proportional to the Doppler factor
$1-\cos{\theta}\,v/c$, where $v$ is the electron velocity and $\theta$
is the angle at which the photon and electron encounter). The other
effect is that photons emerge after scattering preferentially in the
direction of the motion of the relativistic electron. The resulting
angular function contrasts the forward-oriented Klein-Nishina angular
function which corresponds to the case of scattering of high-energy
photons on a resting electron.

The authors certainly realize that the relativistic modification of
the differential cross-section for scattering by an isotropic ensemble
of electrons is implicitly taken into account in any Monte-Carlo simulations of
spectra and angular distributions of photons scattered in mildly or
ultra-relativistic electron gases. In particular, we have verified that results
obtained using the Monte-Carlo Comptonization code of Pozdnyakov et
al. (\cite{pozdnyakov}) and the results of calculations that employ
the formulae of this paper are in perfect agreement.

Apart from a general interest, the formulae obtained here are easily
applicable to the calculations of the scattering albedos (with respect
to low-frequency radiation) of hot atmospheres, such as coronae or
outflows above accretion disks, ADAF accretion flows near black holes,
layers of spreading matter on the surfaces of accreting neutron stars,
intergalactic gas in rich clusters of galaxies, etc. The issue of the
modified albedo may also be important in studying clouds of
ultra-relativistic electrons in compact radiosources, for which an
external radiation may be an additional source of seed photons for
inverse-Compton cooling. Our calculations of albedos demonstrate a
noticeable reduction of the fraction of incident photons which can
propagate inside an optically thick cloud of relativistic
electrons. Therefore, less photons are capable of taking part in the
Comptonization process, and smaller is the rate at which the electrons
lose their energy.

The absence of forward-backward symmetry in the considered regime of
scattering also causes the coefficient of spatial diffusion of photons in
hot plasma to be different from that in the case of non-relativistic
plasma. This affects the shapes of Comptonization spectra and the time delays
between soft and hard radiations coming from variable X-ray sources. A
detailed discussion of both albedo and photon-path-lengthening
consequences of the results reported here is presented in a separate
paper (Sazonov \& Sunyaev \cite{sazonov_a}).

\section{Scattering angular function}

\subsection{Scattering of low-frequency photons in ultra-relativistic plasma}

We start from the formula that describes the differential cross-section for
the Compton scattering by an electron propagating in the direction
$\bg{\omega}$ with a speed $v=\beta c$:
\begin{equation}
\frac{d\sigma}{d\bg{\Omega}^\prime}=\frac{3\sigma_T}{16\pi\gamma^2}\frac{X}
{(1-\beta\bg{\Omega}\bg{\omega})^2}\left(\frac{\nu^\prime}{\nu}\right)^2,
\label{klein}
\end{equation}
where 
\begin{equation}
\frac{\nu^\prime}{\nu}=\frac{1-\beta\bg{\Omega}\bg{\omega}}
{1-\beta\bg{\Omega}^\prime\bg{\omega}+(h\nu/\gamma mc^2)(1-\mu_s)},
\label{en_cons}
\end{equation}
\begin{eqnarray}
X &=& 2-\frac{2(1-\mu_s)}{\gamma^2(1-\beta\bg{\Omega}\bg{\omega})
(1-\beta\bg{\Omega}^\prime\bg{\omega})}
\nonumber\\
&+& \frac{(1-\mu_s)^2}{\gamma^4(1-
\beta\bg{\Omega}\bg{\omega})^2(1-\beta\bg{\Omega}^\prime\bg{\omega})^2}
\nonumber\\
&+& \left(\frac{h\nu}{m_ec^2}\right)^2\frac{\nu^\prime}{\nu}
\frac{(1-\mu_s)^2}{\gamma^2(1-\beta\bg{\Omega}\bg{\omega})^2
(1-\beta\bg{\Omega}^\prime\bg{\omega})^2},
\label{x}
\end{eqnarray}
$\gamma^2=(1-\beta^2)^{-1}$, ($\nu$, $\bg{\Omega}$) are the frequency
and the direction of propagation of the incident photon, ($\nu^\prime$, 
$\bg{\Omega}^\prime$) are the corresponding values for the emergent
photon, $\mu_s=\bg{\Omega}\bg{\Omega}^\prime$, and $\sigma_T$ is the
Thomson scattering cross-section.

Consider now an isotropic distribution of electrons of energy $\gamma m_ec^2$
(which corresponds to a given speed $\beta$). We wish to average
the cross-section given by Eq.~(\ref{klein}) over this velocity
distribution and thereby to find the scattering angular function:
\begin{equation}
p(\gamma,\mu_s)
=\frac{1}{\sigma_T}\int\frac{d\sigma}{d\bg{\Omega}^\prime}
(1-\beta\bg{\Omega}\bg{\omega})d\bg{\omega}.
\label{ang_gen}
\end{equation} 
Here the factor $1-\beta\bg{\Omega}\bg{\omega}$ takes into account the relative
velocity of the photon and electron along the direction of the
latter's motion (Berestetskii et al. \cite{berestetskii}). The factor
$\sigma_T^{-1}$ has been introduced to make $p$ dimensionless and
hence similar to the phase function that is used in the theory of
radiative transfer for describing scattering processes. According to
the definition (\ref{ang_gen}), the mean photon free path can be found as
\begin{equation}
\lambda=\frac{1}{N_e\sigma_T\int p(\gamma,\mu_s)\,d\bg{\Omega}^\prime/4\pi}
=\frac{1}{N_e\sigma_T\int p(\gamma,\mu_s)\,d\mu_s/2},
\label{lambda}
\end{equation}
where $N_e$ is the electron number density.

We shall first consider the limiting case where $\gamma\gg 1$, $\gamma
h\nu/m_ec^2\ll 1$, i.e. the electrons are ultra-relativistic, and the
photons are non-relativistic. The treatment can then be done in the
Thomson limit, and, in a first approximation, the terms proportional
to $h\nu/m_ec^2$ in Eqs. (\ref{en_cons}) and (\ref{x}) can be
ignored. The angular function will therefore be given by the integral
\begin{eqnarray}
p &=& \frac{3}{16\pi}\int\left[\frac{2(1-\beta\bg{\Omega}\bg{\omega})}
{\gamma^2(1-\beta\bg{\Omega}^\prime
\bg{\omega})^2}-\frac{2(1-\mu_s)}{\gamma^4(1-\beta\bg{\Omega}^\prime
\omega)^3}
\right.
\nonumber\\
&&\left.
+\frac{(1-\mu_s)^2}{\gamma^6(1-\beta\bg{\Omega}\bg{\omega})
(1-\beta\bg{\Omega}^\prime\omega)^4}\right] d\bg{\omega}.
\label{ang_3term}
\end{eqnarray}

The main contribution to this integral is provided by the
electrons with $\bg{\omega}$ approaching $\bg{\Omega}^\prime$,
which is the result of Doppler aberration. Therefore, we
may put in Eq.~(\ref{ang_3term}) $\bg{\Omega}\bg{\omega}\approx \mu_s$ and
$1-\beta\bg{\Omega}\bg{\omega}\approx 1-\mu_s$. The following
integral over $\mu^\prime=\bg{\Omega}^\prime\bg{\omega}$ then arises:
\begin{eqnarray}
p &=& \frac{3(1-\mu_s)}{8}\int_{-1}^{1}\left[\frac{2}
{\gamma^2(1-\beta\mu^\prime)^2}-\frac{2}{\gamma^4(1-\beta\mu^\prime)^3}
\right.
\nonumber\\
&&\left.
+\frac{1}{\gamma^6(1-\beta\mu^\prime)^4}\right]d\mu^\prime,
\end{eqnarray}
The integration leads to a very simple result: 
\begin{equation}
p(\gamma\gg 1,\mu_s)=1-\mu_s.
\label{ang_limit}
\end{equation}

The angular function (\ref{ang_limit}) is presented in the left panel
of Fig.~\ref{figure}. It has a simple, apple-like shape (in polar
coordinates), which is the result of the combined action of two
effects, namely the effect of selection of electrons according to the
incidence angle by photons and the Doppler aberration effect, as
discussed in the Introduction.

Using formula (\ref{lambda}), we can find the mean photon free path
that corresponds to the angular function given by
Eq.~(\ref{ang_limit}): $\lambda=(N_e\sigma_T)^{-1}$, as it should be
in the Thomson limit.

It is not difficult to include Klein-Nishina corrections in the expression
(\ref{ang_limit}). We shall do that to the first order of accuracy. We assume
as before that $\gamma h\nu/m_ec^2\ll 1$. In this limit, $h\nu/\gamma
m_ec^2\ll (1-\beta)$, and, therefore, we can expand the ratio
$\nu^\prime/\nu$ given by Eq.~(\ref{en_cons}) as follows:
\begin{equation}
\left(\frac{\nu^\prime}{\nu}\right)^2=\left(\frac{1-\beta\bg{\Omega}
\bg{\omega}}{1-\beta\bg{\Omega}^\prime\bg{\omega}}\right)^2
\left[1-2\frac{h\nu(1-\mu_s)}{\gamma
m_ec^2(1-\beta\bg{\Omega}^\prime\bg{\omega})}+...\right]. 
\end{equation}
The expression in square brackets in the equation above implies that
corrections need to be made to the three terms in Eq.~(\ref{ang_3term}),
namely three additional terms proportional to $h\nu/m_ec^2$ appear. Note
that the fourth term in the expression $X$ (Eq.~[\ref{x}]) leads to a
correction to the angular function of the order of $(\gamma h\nu/ 
m_ec^2)^2$, which will be neglected here. The final result
(the calculation is very similar to the one that had led to
Eq.~[\ref{ang_limit}]) is 
\begin{equation}
p(\gamma,\nu,\mu_s)=1-\mu_s-2\gamma\frac{h\nu}{m_ec^2}(1-\mu_s)^2.
\label{ang_limit1}
\end{equation}

If the electron energy is not too high, say $\gamma\lesssim 10$,
corrections of the form $\gamma^{-n}$ to the angular function become
important. These can be found as follows. On introducing spherical
coordinates with the polar axis pointing along $\bg{\Omega^\prime}$, we obtain
\begin{equation}
\mu=\bg{\Omega}\bg{\omega}=\mu^\prime\mu_s+(1-\mu^{\prime 2})^{1/2}
(1-\mu_s^2)^{1/2}\sin{\phi}. 
\end{equation}
Integration of the first and second bracketed terms in
Eq.~(\ref{ang_3term}) then reduces to 
\begin{equation}
\int\left[\frac{2(1-\beta\mu^\prime\mu_s)}{\gamma^2
(1-\beta\mu^\prime)^2}-\frac{2(1-\mu_s)}{\gamma^4(1-\beta\mu^\prime)^3}
\right]d\mu^\prime,
\label{int_a}
\end{equation}
whereas the third term can be expanded as follows:
\begin{eqnarray}
\int\frac{(1-\mu_s)^2}{\gamma^6(1-\beta\mu^\prime\mu_s)
(1-\beta\mu^\prime)^4}
\nonumber\\
\left[1+\frac{\beta(1-\mu^{\prime 2})^{1/2}(1-\mu_s^2)^{1/2}}
{1-\beta\mu^\prime\mu_s}
\sin{\phi}
\right.
\nonumber\\
\left.
+\frac{\beta^2(1-\mu^{\prime
2})(1-\mu_s^2)}{(1-\beta\mu^\prime\mu_s)^2}\sin^2{\phi}+...\right]
d\mu^\prime d\phi.
\label{int_b}
\end{eqnarray}
Implementing the straightforward integrations in Eqs.~(\ref{int_a})
and (\ref{int_b}) and keeping only the leading terms (of order
$\gamma^{-2}$), we derive
\begin{equation}
p(\gamma,\mu_s)=1-\mu_s+\frac{-1+3\ln{4\gamma^2}}{4\gamma^2}\mu_s.
\label{ang_limit2}
\end{equation} 

We can finally combine Eqs.~(\ref{ang_limit1}) and (\ref{ang_limit2}) to obtain
\begin{equation}
p(\gamma,\nu,\mu_s)=1-\mu_s-2\gamma\frac{h\nu}{m_ec^2}(1-\mu_s)^2
+\frac{-1+3\ln{4\gamma^2}}{4\gamma^2}\mu_s.
\label{ang_limit_corr}
\end{equation}
We implemented a series of Monte-Carlo simulations to determine the
parameter range of applicability of formula (\ref{ang_limit_corr}):
$\gamma\gtrsim 2$, $\gamma h\nu\lesssim 0.02 m_ec^2$. 

In the case of electrons obeying a relativistic Maxwellian
distribution, which is described by the function
$dN_e\propto\gamma(\gamma^2-1)^{1/2}\exp{(-\gamma m_ec^2/kT_e)}\,d\gamma$,
Eq.~(\ref{ang_limit_corr}) should be convolved with this function. The
result is 
\begin{eqnarray}
p(T_e,\nu,\mu_s) =
1-\mu_s-6\frac{h\nu}{m_ec^2}\frac{kT_e}{m_ec^2}(1-\mu_s)^2 
\nonumber\\
+\frac{-1+3\ln{4}
+6\Gamma(0,m_ec^2/kT_e)}{8}\left(\frac{m_ec^2}{kT_e}\right)^2\mu_s,
\label{ang_maxwel1}
\end{eqnarray}
where $T_e$ is the temperature of the electrons and
$\Gamma(\alpha,z)=\int_z^\infty x^{\alpha-1} e^{-x}dx$ is the
incomplete Gamma function. Formula (\ref{ang_maxwel1}) is a good
approximation if $kT_e\gtrsim 2 m_ec^2$, $h\nu kT_e\lesssim 0.01
(m_ec^2)^2$. It is interesting to compare the first of these
conditions with the corresponding constraint on the applicability of
the approximation (\ref{ang_limit_corr}) (see text following
that equation). Evidently, the relatively poor convergence of the
series (\ref{ang_maxwel1}) is due to the contribution of the
low-energy tail of the Maxwellian distribution, i.e. electrons with
$\gamma\lesssim\langle\gamma\rangle= 3kT_e/m_ec^2$.

Various examples of the angular function corresponding to the 
scattering on high-temperature electrons, as resulted from Monte-Carlo
simulations or calculated from Eq.~(\ref{ang_maxwel1}),
are presented in Fig.~\ref{figure}. One can see that Klein-Nishina
corrections, which are described in the first approximation by the
second term on the right-hand side of Eq.~(\ref{ang_maxwel1}),
act to reduce the probability of scattering in all directions. This
reduction reaches a maximum at $\mu_s\rightarrow -1$ (backward
scattering) and monotonically diminishes to become vanishing at
$\mu_s=1$ (forward scattering). The effect of temperature corrections,
which are, to the first order, described by the last term of
Eq.~(\ref{ang_maxwel1}), is that as the temperature decreases, 
progressively more photons become scattered forwards --- see the
pattern corresponding to $kT_e=0.5 m_ec^2$ in the left panel of Figure
\ref{figure} (the result of a Monte-Carlo simulation is shown).

The mean photon free path that corresponds to the angular function
(\ref{ang_maxwel1}), according to Eq.~(\ref{lambda}), is given by
\begin{equation}
\frac{1}{\lambda(T_e,\nu)}=N_e\sigma_T\left(1-8\frac{h\nu}{m_ec^2}
\frac{kT_e}{m_ec^2}\right).
\label{sigma_maxwel1}
\end{equation}
This expression is equivalent to Eq.~(2.17) of Pozdnyakov et
al. (\cite{pozdnyakov}).

In radio galaxies, the relativistic electrons have a
power-law energy spectrum $dN_e\propto\gamma^{-\alpha}d\gamma$ with a
low-energy cutoff $\gamma>\gamma_{min}$. The angular function in this case is
\begin{eqnarray}
p(\alpha,\gamma_{min},\nu,\mu_s) = 1-\mu_s
-\frac{2(\alpha-1)}{\alpha-2}\gamma_{min}\frac{h\nu}{m_ec^2}(1-\mu_s)^2
\nonumber\\
+
\frac{\alpha-1}{4(\alpha+1)}\left(-1+\frac{6}{\alpha+1}+3\ln{4\gamma_{min}^2}
\right)\frac{1}{\gamma_{min}^{2}}\mu_s.
\end{eqnarray}
This formula is applicable if $\gamma_{min}\gg 1$, $\alpha>2$ and
$\gamma_{min}h\nu/m_ec^2\ll 1$.

\subsection{Scattering of mildly relativistic photons in mildly
relativistic thermal plasma ($h\nu,kT_e\ll m_ec^2$)}

In this case, the angular function can be approximated by the
following formula, which was derived earlier in (Sazonov \& Sunyaev
\cite{sazonov_b}),
\begin{eqnarray}
p(T_e,\nu,\mu_s)=\frac{3}{4}\left[1+\mu_s^2-2(1-\mu_s)(1+\mu_s^2)
\frac{h\nu}{m_ec^2}
\right.
\nonumber\\
\left.
+2(1-2\mu_s-3\mu_s^2+2\mu_s^3)\frac{kT_e}{m_ec^2}
\right.
\nonumber\\
\left.
+(1-\mu_s)^2(4+3\mu_s^2)\left(\frac{h\nu}{m_ec^2}\right)^2
\right.
\nonumber\\
\left.
+
(1-\mu_s)(-7+14\mu_s+9\mu_s^2-10\mu_s^3)\frac{h\nu}{m_ec^2}\frac{kT_e}{m_ec^2}
\right.
\nonumber\\
\left.
+(-7+22\mu_s+9\mu_s^2-38\mu_s^3+20\mu_s^4)\left(\frac{kT_e}{m_ec^2}\right)^2
+...\right].
\label{ang_maxwel2}
\end{eqnarray}
The main term in this power series, $3(1+\mu_s^2)/4$, is the usual
Rayleigh function, which corresponds to the non-relativistic case. 

The range of applicability of formula (\ref{ang_maxwel2}) is roughly
$kT_e\lesssim 0.05 m_ec^2$ and $h\nu\lesssim 0.1 m_ec^2$. Some
examples are displayed in the right panel of Fig.~\ref{figure}. The effect of
the first-order temperature correction (the term $\propto
kT_e/m_ec^2$ in Eq. [\ref{ang_maxwel2}]) on the angular function is to
enhance the number of photons scattered at intermediate angles
between $69^\circ$ and $138^\circ$ (a maximum of $12 kT_e/(0.05
m_ec^2)$ per cent is reached at an angle of $105^\circ$) and to
suppress scattering in both forward and backward directions (by $10
kT_e/(0.05 m_ec^2)$ per cent at angles 0 and $\pi$). 

The mean photon free path corresponding to the angular function
(\ref{ang_maxwel2}) is given by 
\begin{eqnarray}
\frac{1}{\lambda(T_e,\nu)}=N_e\sigma_T
\left[1-2\frac{h\nu}{m_ec^2}-5\frac{h\nu}{m_ec^2}\frac{kT_e}{m_ec^2}
\right.
\nonumber\\
\left.
+\frac{26}{5}\left(\frac{h\nu}{m_ec^2}\right)^2\right],
\label{sigma_maxwel2}
\end{eqnarray}
which is a well-known expression (see, e.g., Pozdnyakov et
al. \cite{pozdnyakov}).

\begin{acknowledgements}
This work was supported in part by the Russian Foundation for Basic
Research through grant 97-02-16264.
\end{acknowledgements}

\end{document}